\documentclass[pre,reprint,amssymb]{revtex4-1}
\usepackage[english]{babel}
\usepackage{amsmath}
\usepackage{amsfonts}
\usepackage{amssymb}
\usepackage{array}
\usepackage{dcolumn}
\usepackage{longtable}
\usepackage{hyperref}
\usepackage{float}
\usepackage{bbm}
\usepackage{bm}
\usepackage{graphicx}
\usepackage{natbib}
\usepackage{verbatim}

\def\d{\text{d}}

\newcommand{\eins}{\mbox{$1 \hspace{-1.0mm} {\bf l}$}}

\begin{document}
\title{Requirements for contractility in disordered cytoskeletal bundles}
\author{Martin Lenz}
\email{martinlenz@uchicago.edu}
\affiliation{James Franck Institute, University of Chicago, Chicago IL 60637, USA}
\author{Margaret L. Gardel}
\affiliation{James Franck Institute, University of Chicago, Chicago IL 60637, USA\\
Institute for Biophysical Dynamics, University of Chicago, Chicago IL 60637, USA\\
Department of Physics, University of Chicago, Chicago IL 60637, USA}
\author{Aaron R. Dinner}
\affiliation{James Franck Institute, University of Chicago, Chicago IL 60637, USA\\
Institute for Biophysical Dynamics, University of Chicago, Chicago IL 60637, USA\\
Department of Chemistry, University of Chicago, Chicago IL 60637, USA}

\begin{abstract}
Actomyosin contractility is essential for biological force generation, and is well understood in highly organized structures such as striated muscle. Additionally, actomyosin bundles devoid of this organization are known to contract both \emph{in vivo} and \emph{in vitro}, which cannot be described by standard muscle models. To narrow down the search for possible contraction mechanisms in these systems, we investigate their microscopic symmetries. We show that contractile behavior requires non-identical motors that generate large enough forces to probe the nonlinear elastic behavior of F-actin. This suggests a role for filament buckling in the contraction of these bundles, consistent with recent experimental results on reconstituted actomyosin bundles.
\end{abstract}


\maketitle

\section{\label{sec:Introduction}Introduction}
\begin{figure}[b]
\resizebox{8.5cm}{!}{\includegraphics{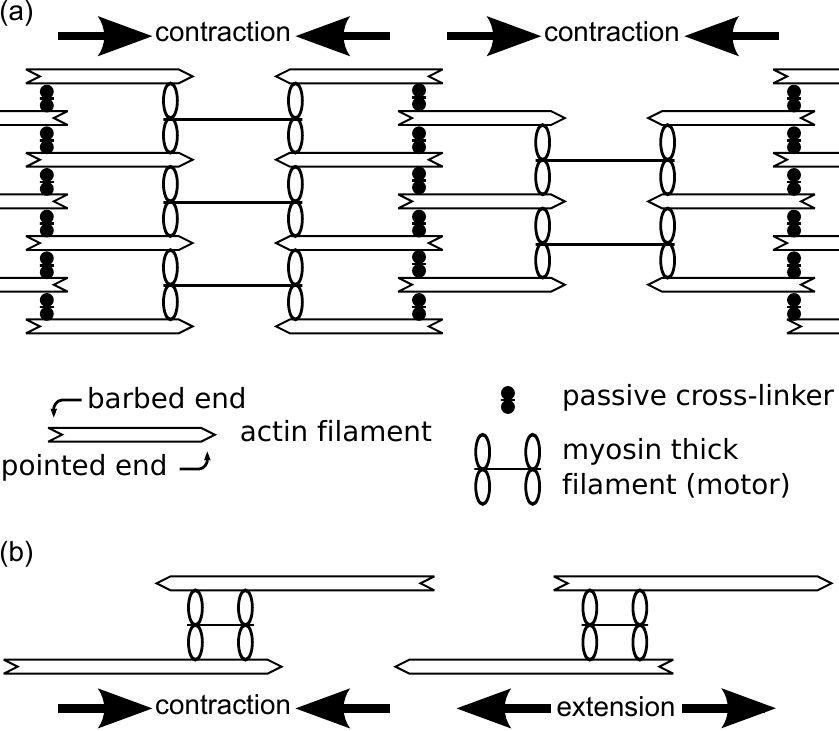}}
\caption{\label{fig:muscle}Contraction and extension in actomyosin systems. (a)~Contraction in sarcomeres occurs as motor localized in the vicinity of the F-actin pointed ends slide towards F-actin barbed ends.
(b)~A motor located near the F-actin pointed ends induces local contraction as in sarcomeres (\emph{left}), while localization near the barbed ends yields extension (\emph{right}). The two effects balance each other in a large class of bundles, which we characterize in this paper.
}
\end{figure}

The large-scale motion of living organisms often depends on their ability to harness the power of nanometer-sized molecular motors to generate macroscopic displacements. For instance, our striated muscles rely on clusters---or ``thick filaments''---of the molecular motor myosin to generate forces. Myosin thick filaments are able to slide directionally towards the barbed end of polar actin filaments (F-actin), and the characteristic organization of striated muscles into periodic sarcomeres arranged in series allows the transfer of this microscopic motion to larger scales [Fig.~\ref{fig:muscle}(a)] \cite{Alberts:1998aa}.

Despite its familiarity, sarcomere-like organization is far from a universal feature of contractile actomyosin assemblies. In some instances, partially periodic arrangements reminiscent of sarcomeres are observed, as in subcellular contractile bundles known as stress fibers \cite{Peterson:2004}. In many other cases, however, no such organization is known to exist. Examples include smooth muscle fibers \cite{Fay:1983}, transverse arcs \cite{Heath:1983}, graded polarity bundles \cite{Cramer:1997}, the cell cortex \cite{Medalia:2002} and lamellar networks \cite{Verkhovsky:1995}. Sarcomere-like contraction is unlikely to apply to these systems, and there is no consensus regarding their actual deformation mechanism.

\emph{In vitro} experiments using purified proteins are useful for understanding contraction in these systems,
and have been used to identify the minimum requirements of actomyosin contractility since the 1940s \cite{Szent-Gyorgyi:1947}. Modern
attempts using dilute actomyosin gels were not able to induce contractility in the presence of actin and myosin alone, although adding the actin cross-linker $\alpha$-actinin did produce observable contraction \cite{Janson:1991,Mizuno:2007aa,Bendix:2008,Koenderink:2009}. However, a more recent study using denser actomyosin bundles shows that F-actin and myosin can induce contractility on their own \cite{Thoresen:2011}. Unlike in sarcomeres, in these bundles F-actin lacks polarity ordering and myosins are not aligned in register, and we thus refer to them as ``disordered''. 

Previous theoretical work on disordered actomyosin systems include continuum models focused on length scales much larger than an individual actin filament \cite{Kruse:2003ab,Kruse:2004aa,Kruse:2005aa,MacKintosh:2008aa}. In these elegant descriptions, contractility is introduced phenomenologically, which circumvents the question of its emergence from microscopic interactions. Several other studies do however investigate this connection. In simulations without sarcomeric organization or cross-linkers, thick filaments simply sort F-actin by polarity without inducing any overall contraction \cite{Zemel:2009}. To restore contractility, several models assume that thick filaments tend to dwell at the barbed end of F-actin after sliding over its whole length \cite{Kruse:2000,Kruse:2003aa,Liverpool:2003,Ziebert:2005}, or more generally that their velocity depends on their position relative to the filament \cite{Liverpool:2005}. In these models, F-actin tend to have immobilized motors that transiently act as passive cross-linkers at their barbed ends. This essentially introduces a small amount of sarcomere-like organization and results in contractility \cite{Zemel:2009}. However, no direct experimental evidence of thick filaments dwelling at the barbed end of F-actin is available.

Here we investigate the possibility of bundle contraction in the absence of any sarcomeric organization, including motors dwelling at the filament barbed ends. After introducing a general bundle model in Sec.~\ref{sec:Bundle Model}, we show in Sec.~\ref{sec:Situations without telescopic deformation} that underlying symmetries between contraction and extension [Fig.~\ref{fig:muscle}(b)] imply that disordered bundle contraction requires non-identical motors and a non-linear elastic behavior of the filaments. 
Intuitively, non-identical motors induce mechanical frustration in a disordered actomyosin bundle, thus generating both contractile and extensile stresses. The filament non-linear behavior then allows the former to deform the bundle while resisting the latter, yielding overall contraction.
Finally, in Sec.~\ref{sec:Discussion} we discuss the limitations of our model and propose a minimal model for the contraction of
a non-sarcomeric bundle.

\section{\label{sec:Bundle Model}Bundle model} 
In order to make general statements about 
a disordered bundle of potentially complex internal geometry, we develop a detailed description of its
mechanics without resorting to the simplifying mean-field approximations widely used in previous studies \cite{Kruse:2000,Kruse:2003aa,Liverpool:2003,Ziebert:2005,Liverpool:2005}. The main assumptions of our model are as follows. First, we assume that the velocity of a motor only depends on the force applied to it by the F-actin to which it is bound. Second, the average deformation of a thermally fluctuating section of filament in the bundle only depends on the force applied longitudinally at its ends. Third, we consider
a stabilized bundle
where F-actin polymerization and depolymerization do not occur. Finally, our model does not include the attachment-detachment dynamics of motors and cross-linkers to F-actin. The relevance of these choices is discussed in Sec.~\ref{sec:Discussion}.

We describe
a bundle of arbitrary geometry by subdividing it
into interacting ``units'' of three types characterized by their lengths, velocities and tensions (Sec.~\ref{sec:Linkers, filaments and junctions}). Introducing general notations to describe the physical connections between different units, we express the relationships between their forces and velocities as a function of their spatial arrangement (Sec.~\ref{sec:Bundle geometry and topology}). The central physics of bundle mechanics are then described by introducing the filament force-extension relationships and the motor force-velocity relationships (Sec.~\ref{sec:Filament elasticity and motor operation}). Conservation equations are then used to derive a compact description of the bundle amenable to further discussion (Sec.~\ref{sec:Dynamical equations for bundle deformation}).

\subsection{\label{sec:Linkers, filaments and junctions}Linkers, filaments and junctions}
\begin{figure}[t]
\resizebox{8.5cm}{!}{\includegraphics{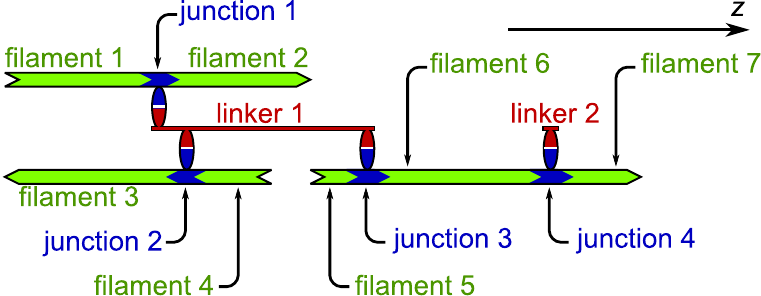}}
\caption{\label{fig:bundle}
(Color online) Schematic of a model bundle comprising three F-actin, one myosin thick filament bound in three different sites (linker~1) and one bound only once (linker~2). Colors indicate the division into linker units (\emph{red}), junction units (\emph{blue}) and filament units (\emph{green}). Our model does not impose any restrictions on the number of junctions associated with a linker, whether they involve one or more F-actin, or the polarity of the F-actin.
}
\end{figure}

Here we consider a single bundle
constituted of F-actin, thick filaments of myosin molecular motors, and optionally passive actin cross-linkers. F-actin are assumed to be aligned in the $z$ direction, and for simplicity we refer to the direction of positive (negative) $z$ as the ``right'' (``left'') in the following. To describe the bundle, we decompose it so as to distinguish three types of ``units'' (Fig.~\ref{fig:bundle}):
\begin{itemize}
\item\textbf{linker units}, representing a whole myosin thick filament or an passive actin cross-linker (passive cross-linkers are equivalent to immobile motors). The total number of linker units in the system is denoted as $n''$.
\item\textbf{junction units}, representing the point of contact between a myosin thick filament or passive actin cross-linker on the one hand, and an F-actin on the other. The total number of junction units in the system is denoted as $n'$.
\item\textbf{filament units}, representing a \emph{portion} of an F-actin comprised between two junction units, or between one junction unit and an F-actin free end. We do not consider freely floating F-actin (filament units with two free ends). The total number of filament units in the system is denoted as $n$.
\end{itemize}

\begin{figure}[t]
\resizebox{8.5cm}{!}{\includegraphics{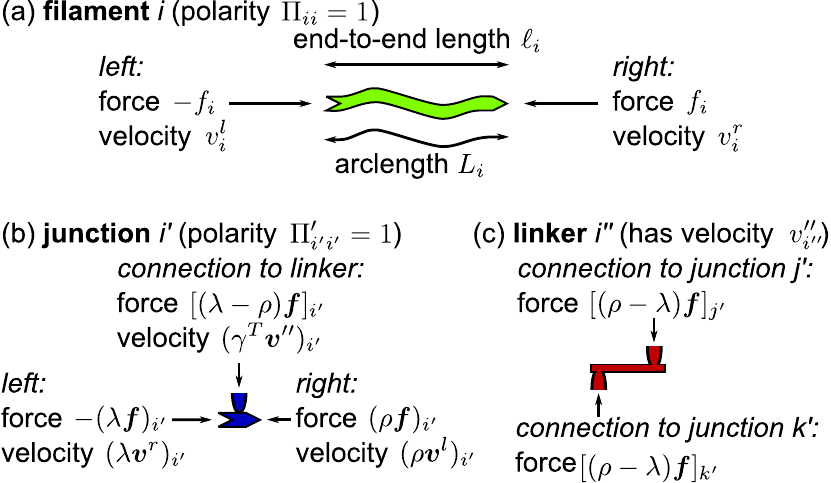}}
\caption{\label{fig:notation}
(Color online) Summary of notation. A filament unit (a), junction unit (b), and linker unit (c) are represented, along with the velocities and lengths characterizing their state, as well as the forces they are subjected to.
}
\end{figure}

We label the filament units by $i=1,\ldots,n$, and denote by $f_i$ the tension of filament unit $i$ ($f_i>0$ for a filament unit under extension). We allow filament units to bend away from the $z$-axis while maintaining their overall $z$ orientation, and thus introduce filament unit $i$'s contour length and end-to-end length as two independent variables $L_i$ and $\ell_i$, respectively. Finally, we introduce the velocity $v_i^r$ of the rightmost monomer of filament unit $i$, and the velocity $v_i^l$ of its leftmost monomer. 
These two velocities are Eulerian velocities; in that respect, one might think of junction 1 of Fig.~\ref{fig:bundle} as a bridge and of the actin as the water that flows under it. Then the velocities $v_1^r$ and $v_2^l$ associated with filament units 1 and 2 are the velocities of the water just upstream and downstream from the bridge, as opposed to being the velocities of a specific fluid element.
The notation is summarized in Fig.~\ref{fig:notation}(a).

We label junction units by $i'=1,\ldots,n'$, and linker units by $i''=1,\ldots,n''$. We introduce the velocity $v_{i''}''$ of linker unit $i''$,
\emph{i.e.}, the velocity of the bridge itself in our previous analogy.
We choose to work in the reference frame where the center-of-mass of all linker units is motionless, which reads
\begin{equation}\label{eq:referenceframe}
\sum_{i''=1}^{n''}v''_{i''}=0.
\end{equation}
Note that the velocities $v_i^r$ and $v_i^l$ are absolute velocities defined in this reference frame, as opposed to being velocities relative to the neighboring linker.
We delay labeling the other forces and velocities involved in junction and linker units until the introduction of convenient notations in the next section.

\subsection{\label{sec:Bundle geometry and topology}Bundle geometry and topology}
To describe the physical connections between junction and filament units, we define the $n'\times {n}$ matrices $\rho$ and $\lambda$ by
\begin{subequations}
\begin{eqnarray}
\rho_{i'i}&=&
\begin{cases} 1 & \text{if $i$ is the right-hand neighbor of $i'$}
\\
0 & \text{otherwise}
\end{cases}\\
\lambda_{i'i}&=&
\begin{cases} 1 & \text{if $i$ is the left-hand neighbor of $i'$}
\\
0 & \text{otherwise.}
\end{cases}
\end{eqnarray}
\end{subequations}
For instance, the bundle represented in Fig.~\ref{fig:bundle} is described by 
\begin{equation}\label{eq:lambdarhoexample}
\lambda=\left(
\begin{array}{ccccccc}
1 & 0 & 0 & 0 & 0 & 0 & 0\\
0 & 0 & 1 & 0 & 0 & 0 & 0\\
0 & 0 & 0 & 0 & 1 & 0 & 0\\
0 & 0 & 0 & 0 & 0 & 1 & 0
\end{array}
\right),
~
\rho=\left(
\begin{array}{ccccccc}
0 & 1 & 0 & 0 & 0 & 0 & 0\\
0 & 0 & 0 & 1 & 0 & 0 & 0\\
0 & 0 & 0 & 0 & 0 & 1 & 0\\
0 & 0 & 0 & 0 & 0 & 0 & 1
\end{array}
\right).
\end{equation}
The usefulness of these matrices is illustrated by introducing the notation $\bm{f}=(f_1,\ldots,f_{{n}})$ for the vector of all filament tensions, as well as similar notations $\bm{L}$, $\bm{\ell}$, $\bm{v}^r$, $\bm{v}^l$, $\bm{v}''$. The matrix product $\rho\bm{f}$ is a vector of length $n'$ whose $i'$th component $(\rho\bm{f})_{i'}$ is the tension of the filament unit that is the right-hand neighbor of junction unit $i'$. This neighbor thus exerts a force $(\rho\bm{f})_{i'}$ on junction unit $i'$, while its left-hand neighbor exerts $-(\lambda\bm{f})_{i'}$. The sum of these two forces is equal and opposite to the force applied to junction unit $i'$ by its linker unit. This last force thus reads $[(\lambda-\rho)\bm{f}]_{i'}$ [Fig.~\ref{fig:notation}(b)].

Further use of $\rho$ and $\lambda$ indicates that the velocities of the rightmost and leftmost actin monomers involved in junction unit $i'$ are $(\rho\bm{v}^l)_{i'}$ and $(\lambda\bm{v}^r)_{i'}$, respectively [Fig.~\ref{fig:notation}(b)]. As junction units are point-like objects, the net actin flow in and out of them vanishes and
\begin{equation}\label{eq:filamentvelocitycontinuity}
\lambda\bm{v}^r=\rho\bm{v}^l.
\end{equation}

The reasonings used here can be generalized in the following way: if $x_i$ is a quantity associated with filament $i$, then $(\rho\bm{x})_{i'}$ is associated to the right-hand neighbor of junction unit $i'$. For instance, in the two previous paragraphs we considered $x_i=f_i$ and $x_i=v_i^l$ or $v_i^r$, respectively. Just as this statement relates junction units to quantities associated with the neighboring filament units, we can conversely relate filament units to quantities associated with their junction unit neighbors as follows. If $x'_{i'}$ is associated with junction unit $i'$, then $(\rho^T\bm{x}')_{i}$ is associated with the junction unit that is the \emph{left} neighbor of filament unit $i$ if it has one, or is equal to zero if it does not (the superscript $T$ denotes the matrix transpose). A similar statement holds for $\lambda^T$. Combining this with the fact that each junction unit has exactly one right-hand and one left-hand neighbor, we find that $(\rho\rho^T\bm{x}')_{i'}=x'_{i'}$ always. More generally,
\begin{equation}
\rho\rho^T=\lambda\lambda^T=\eins,
\end{equation}
where $\eins$ denotes the identity matrix. We further note that filament units may have one or two neighbors, which implies
\begin{equation}\label{eq:projector}
(\rho^T\rho\bm{x})_{i}=
\begin{cases} x_i & \text{if $i$ has a left-hand neighbor}
\\
0 & \text{if it does not.}
\end{cases}\\
\end{equation}
This means that matrix $\rho^T\rho$ is a projector onto the subspace of filament units that have a left-hand neighbor. A similar statement holds for $\lambda^T\lambda$. This discussion implies that $n'<n$.

To describe the polarity of the filament units, we introduce the diagonal $n\times n$ matrix $\Pi$, where $\Pi_{ii}=1$ if the pointed end of filament unit $i$ points to the right, and $\Pi_{ii}=-1$ if it points to the left. As an example, Fig.~\ref{fig:bundle} has 
\begin{equation}\label{eq:Piexample}
\Pi=
\textrm{diag}(1,1,-1,-1,1,1,1)
\end{equation}
A similar diagonal polarity matrix $\Pi'$ is associated with junction units, and since neighboring filaments and junction units have the same polarity we have
\begin{equation}\label{eq:pprime}
\Pi'=\rho \Pi\rho^T=\lambda \Pi\lambda^T.
\end{equation}

Turning to the linker units, we define the $n''\times n'$ matrix $\gamma$ by
\begin{equation}
\gamma_{i''i'}=
\begin{cases} 1 & \text{if $i'$ is connected to $i''$}
\\
0 & \text{otherwise.}
\end{cases}
\end{equation}
For instance, Fig.~\ref{fig:bundle} has
\begin{equation}\label{eq:gammaexample}
\gamma=\left(
\begin{array}{cccc}
1 & 1 & 1 & 0\\
0 & 0 & 0 & 1\\
\end{array}
\right).
\end{equation}
Therefore, the velocity of the linker unit connected to junction unit $i'$ is $(\gamma^T\bm{v}'')_{i'}$ [Fig.~\ref{fig:notation}(b)]. Each junction unit is connected to a linker unit, but one linker unit can be connected to several junction units, which implies $n''\leqslant n'$.

To obtain the forces associated with the linker units, we reason that if $i'$ experiences a force $[(\lambda-\rho)\bm{f}]_{i'}$ from its linker unit, then the $i'$ exerts an equal and opposite force $[(\rho-\lambda)\bm{f}]_{i'}$ on the linker unit [Fig.~\ref{fig:notation}(c)]. Force balance imposes that the sum of the forces applied on any linker unit vanishes, and thus
\begin{equation}
\sum_{i'\text{ connected to }i''}[(\rho-\lambda)\bm{f}]_{i'}=0,
\end{equation}
or, in vector notation:
\begin{equation}\label{eq:linkerforce}
\gamma(\rho-\lambda)\bm{f}=0.
\end{equation}
Although this condition comprises $n''$ scalar equations, those equations are not all independent. This can be seen by considering the mechanical subsystem formed by all junction and filament units (but not including the linker units). As inertia and friction against the background fluid are negligible, the sum of all forces applied to this system by junction units must vanish:
\begin{equation}\label{eq:subsystemforcebalance}
\sum_{i'=1}^{n'}[(\lambda-\rho)\bm{f}]_{i'}=-\sum_{i''= 1}^{n''}[\gamma(\rho-\lambda)\bm{f}]_{i''}=0,
\end{equation}
where the first equality follows from the fact that $\gamma$ has exactly one element equal to 1 per column and zeros everywhere else. As Eq.~(\ref{eq:subsystemforcebalance}) is always trivially true, Eq.~(\ref{eq:linkerforce}) expresses only $n''-1$ linearly independent scalar conditions.

\subsection{\label{sec:Filament elasticity and motor operation}Filament elasticity and motor operation}
Having defined notations for all lengths, forces and velocities in our system (Fig.~\ref{fig:notation}), as well as enforced velocity continuity and force balance conditions, we turn to characterizing the more substantial physics of the junction and filament units.

Filament units are sections of semiflexible polymers shorter than or with a length comparable to their persistence length.
We thus assume that the force required to hold a filament unit of given contour length $L_i$ in mechanical equilibrium is uniquely determined by specifying its end-to-end length $\ell_i$, which defines the force-extension relationship $F$:
\begin{equation}\label{eq:generalfext}
\quad f_i=F(\ell_i,L_i).
\end{equation}
In the case of a thermally fluctuating polymer, $\ell_i$ denotes the end-to-end length averaged over thermal fluctuations. Thermal bending of the filament unit moreover implies that $\ell_i$ is smaller than $L_i$ even when $f_i=0$. Since filament units are shorter than the filament persistence length, we expect deformations of this kind ranging from zero to $\approx 20\%$.
In vector notation, we write
\begin{equation}\label{eq:vectorfext}
\bm{f}=\bm{F}(\bm{\ell},\bm{L}),
\end{equation}

Motor operation at junction $i'$ is described by a functional relationship between the local velocity of the linker relative to the F-actin and the force applied to $i'$:
\begin{equation}\label{eq:nopolfv}
(\rho\bm{v}^l-\gamma^T\bm{v}'')_{i'}
=
\tilde{V}'_{i'}\left\lbrace
[(\lambda-\rho)\bm{f}]_{i'}
\right\rbrace,
\end{equation}
where the function $\tilde{V}'_{i'}$ \emph{a priori} depends on the polarity $\Pi'_{i'i'}$ of the junction unit. This dependence can be explicitly determined by noting that the force-velocity relationship must not depend on our arbitrary choice of the direction of positive $z$. Reversing this choice is equivalent to reversing the sign of all velocities, forces and polarities. The only way for Eq.~(\ref{eq:nopolfv}) to be invariant under this transformation is to write
\begin{equation}
(\rho\bm{v}^l-\gamma^T\bm{v}'')_{i'}
=\Pi'_{i'i'}V'_{i'}\left\lbrace
\Pi'_{i'i'}[(\lambda-\rho)\bm{f}]_{i'}
\right\rbrace,
\end{equation}
where function $V'_{i'}$ is the force-velocity relationship of motor $i'$, an \emph{a priori} nonlinear function independent of $\Pi'$. In vector notation,
\begin{equation}\label{eq:fv}
\rho\bm{v}^l-\gamma^T\bm{v}''
=\Pi'\bm{V}'\left[
\Pi'(\lambda-\rho)\bm{f}
\right].
\end{equation}

\subsection{\label{sec:Dynamical equations for bundle deformation}Dynamical equations for bundle deformation}
To provide a kinematic description of bundle contraction and extension, we write the conservation of F-actin contour length. The rate of change of the contour length of a filament unit is directly related to the velocity at which the neighboring junction units slide relative to F-actin. If $i$ has a neighboring junction unit on its right-hand side, the velocity of the linker unit there is $(\lambda^T\gamma^T\bm{v}'')_i$, and its sliding velocity relative to the actin is $(\lambda^T\gamma^T\bm{v}''-\bm{v}^r)_i$. We rewrite this sliding velocity as $(\lambda^T\gamma^T\bm{v}''-\lambda^T\lambda\bm{v}^r)_i$, which is equal to the previous expression if $i$ has a right-hand neighbor, and to zero if it does not. Using a similar reasoning for the left-hand side, we obtain
\begin{equation}\label{eq:darclengthdt}
\frac{\d \bm{L}}{\d t}=(\lambda^T\gamma^T\bm{v}''-\lambda^T\lambda\bm{v}^r)-(\rho^T\gamma^T\bm{v}''-\rho^T\rho\bm{v}^l).
\end{equation}
The first (second) term on the right-hand side of this equation accounts for actin-linker sliding on the right-hand (left-hand) side of the filaments units, and vanishes for the filaments units that do not have a right-hand (left-hand) neighbor junction unit.

We describe the evolution of the end-to-end length of a filament unit in different ways depending on whether it has two junction unit neighbors or has a free end. In the former case,
\begin{equation}
\frac{\d \ell_i}{\d t}=\left(\lambda^T\gamma^T\bm{v}''\right)_i-\left(\rho^T\gamma^T\bm{v}''\right)_i,
\end{equation}
where the two terms in the right-hand side are the absolute velocities of the right and left neighbors of $i$, respectively. To describe a filament unit with one free end, we first note that it has vanishing tension. If the contour length of the filament unit is known, its end-to-end length is given by its force-extension relation Eq.~(\ref{eq:generalfext}) with $f_i=0$. Differentiating with respect to time, we obtain
\begin{equation}\label{eq:differentiatefloppiness}
\frac{\partial F}{\partial \ell}\frac{\d \ell_i}{\d t}+\frac{\partial F}{\partial L}\frac{\d L_i}{\d t}=0.
\end{equation}
Defining the diagonal matrices
\begin{subequations}
\begin{eqnarray}
\left(\frac{\partial\bm{F}}{\partial\bm{L}}\right)_{ij}
=\frac{\partial f_i}{\partial L_j}
&=&\delta_{ij}\frac{\partial F}{\partial L}(\ell_i,L_i)\\
\left(\frac{\partial\bm{F}}{\partial\bm{\ell}}\right)_{ij}
=\frac{\partial f_i}{\partial\ell_j}
&=&\delta_{ij}\frac{\partial F}{\partial \ell}(\ell_i,L_i),
\end{eqnarray}
\end{subequations}
where $\delta_{ij}$ is the Kronecker delta, we rewrite Eq.~(\ref{eq:differentiatefloppiness}) as
\begin{equation}
\frac{\d \ell_i}{\d t}=-\left[\left(\frac{\partial \bm{F}}{\partial \bm{\ell}}\right)^{-1}\frac{\partial \bm{F}}{\partial \bm{L}}\frac{\d \bm{L}}{\d t}\right]_i
\end{equation}
for a filament unit with a free end \footnote{Mechanical stability imposes that ${\partial{F}}/{\partial{\ell}}$ is always strictly negative, making the matrix ${\partial\bm{F}}/{\partial\bm{\ell}}$ invertible.}. Making use of Eq.~(\ref{eq:projector}) and its analog for $\lambda^T\lambda$, we write an equation that describes filament units whether they have one or two neighboring junction units:
\begin{eqnarray}
\frac{\d \bm{\ell}}{\d t}&=&(\rho^T\rho\lambda^T-\lambda^T\lambda\rho^T)\gamma^T\bm{v}''\nonumber\\
&&-(\eins-\rho^T\rho\lambda^T\lambda)\left(\frac{\partial\bm{F}}{\partial\bm{\ell}}\right)^{-1}\frac{\partial\bm{F}}{\partial\bm{L}}\frac{\d \bm{L}}{\d t}.\label{eq:dlengthdt}
\end{eqnarray}

We finally use Eqs.~(\ref{eq:filamentvelocitycontinuity}), (\ref{eq:vectorfext}) and (\ref{eq:fv}) to eliminate $\bm{f}$, $\rho\bm{v}^l$ and $\lambda\bm{v}^r$ in Eqs.~(\ref{eq:darclengthdt}) and (\ref{eq:dlengthdt}). This yields
\begin{widetext}
\begin{subequations}\label{eq:diffsystem}
\begin{eqnarray}
\frac{\d\bm{L}}{\d t}
&=&(\rho^T-\lambda^T)\Pi'
\bm{V}'[\Pi'(\lambda-\rho)\bm{F}(\bm{\ell},\bm{L})]\label{eq:diffsystemL}
\\
\frac{\d\bm{\ell}}{\d t}
&=&
-\left(\frac{\partial\bm{F}}{\partial\bm{\ell}}\right)^{-1}\frac{\partial\bm{F}}{\partial\bm{L}}
(\rho^T\rho-\lambda^T\lambda)(\rho^T+\lambda^T)\Pi'
\bm{V}'[\Pi'(\lambda-\rho)\bm{F}(\bm{\ell},\bm{L})]
+(\rho^T\rho\lambda^T-\lambda^T\lambda\rho^T)\gamma^T\bm{v}''_s(\bm{\ell},\bm{L}),\label{eq:diffsystemell}
\end{eqnarray}
\end{subequations}
where the nonlinear vector function $\bm{v}''_s(\bm{\ell},\bm{L})$ is the solution of the linear (in $\bm{v}''$) system of equations formed by Eq.~(\ref{eq:referenceframe}) and the following vector equation, obtained by combining Eqs.~(\ref{eq:linkerforce}) and (\ref{eq:vectorfext}), then differentiating with respect to time and inserting Eqs.~(\ref{eq:diffsystem}) into the result:
\begin{equation}\label{eq:v''eq}
\gamma(\rho-\lambda)\frac{\partial\bm{F}}{\partial\bm{\ell}}(\rho^T\rho\lambda^T-\lambda^T\lambda\rho^T)\gamma^T\bm{v}''
=
\gamma(\rho-\lambda)\frac{\partial\bm{F}}{\partial\bm{L}}(\rho^T\rho\lambda^T-\lambda^T\lambda\rho^T)
\Pi'\bm{V}'[\Pi'(\lambda-\rho)\bm{F}(\bm{\ell},\bm{L})].
\end{equation}
\end{widetext}
Equation (\ref{eq:v''eq}), just like Eq.~(\ref{eq:linkerforce}), has only $n''-1$ independent scalar equations, and supplementing it with Eq.~(\ref{eq:referenceframe}) thus results in a complete set of equations for $\bm{v}''_s$. Finally, supplementing Eqs.~(\ref{eq:referenceframe}), (\ref{eq:diffsystem}) and (\ref{eq:v''eq}) with an initial condition $[\bm{\ell}(t=0),\bm{L}(t=0)]$ completely specifies the dynamics of a bundle of arbitrary geometry.

\section{\label{sec:Situations without telescopic deformation}Situations without telescopic deformation}
We now ask under what conditions contraction occurs. 
We are interested in bundles much longer than the size of any single one of their constituents (F-actin or motor). Significant contraction of such a bundle requires that it contracts throughout its length, as opposed to, \emph{e.g.}, contracting at its extremities while the bulk of the bundle retains a constant length. We thus
focus on ``telescopic deformation'', whereby the end-to-end velocity of a bundle contracting under vanishing external load is proportional to its length. This constitutes the standard behavior of contractile actomyosin structures
\emph{in vivo} \cite{Alberts:1998aa,Bement:1991,Herrera:2005,Carvalho:2009} and \emph{in vitro} \cite{Thoresen:2011}. Telescopic deformation
is characteristic of systems formed of a serial arrangement of many independently deforming elements, often referred to as ``contractile units'' \cite{Thoresen:2011,Herrera:2005,Carvalho:2009}.

Here we demonstrate two requirements for telescopic deformation. In Sec.~\ref{sec:Motors with identical force-velocity relationships} we show that it cannot arise if the motors all have identical force-velocity relationships. Sec.~\ref{sec:Disordered bundles with linearly elastic filaments} then tackles situations where motors with different force-velocity relationships are present. In that case, we prove that bundles lacking polarity organization (\emph{e.g.}, sarcomeres) comprised of linearly elastic (\emph{i.e.}, rigid) filaments do not undergo telescopic deformation.
Such rigid filaments units represent situations where the filament persistence length is very large (\emph{e.g.}, bundles of microtubules and kinesin oligomers), or when the filament units themselves are very short (\emph{e.g.}, strongly cross-linked bundles, where the spacing between two junctions is small).

To completely determine the bundle dynamics, we need to specify an initial condition $[\bm{\ell}(t=0),\bm{L}(t=0)]$. We thus choose an arbitrary vector $\bm{L}_0$ of length $n$ and impose $\bm{L}(t=0)=\bm{L}_0$.
To avoid confusion between the effects of motor-generated stresses, which are relevant for contractility, and those of bundle prestress, which are not, we consider bundles that are initially stress-free:
\begin{equation}\label{eq:noprestress}
\bm{f}(t=0)=\bm{F}[\bm{\ell}(t=0),\bm{L}(t=0)]=0.
\end{equation}
This imposes $\bm{\ell}(t=0)=\bm{\ell}_0$, where $\bm{\ell}_0$ is the solution of the equation $\bm{F}(\bm{\ell}_0,\bm{L}_0)=0$, and thus completely specifies the bundle initial condition.

\subsection{\label{sec:Motors with identical force-velocity relationships}Motors with identical force-velocity relationships}
In a bundle where the motors have identical force-velocity relationships, the junction units have identical spontaneous sliding velocities $v^*$ in the absence of applied force. Defining $\bm{v}^*$ as the length $n'$ vector with all its components equal to $v^*$, we can thus write
\begin{equation}\label{eq:identicalunloaded}
\bm{V}'(\bm{f}=0)=\bm{v}^*.
\end{equation}
We now demonstrate that under this assumption, all linker units are immobile throughout the dynamics and all right-pointing filaments undergo a uniform translation with constant velocity $v^*$, while left-pointing filaments translate with $-v^*$.

We define the set of functions $[\bm{\ell}^*(t),\bm{L}^*(t)]$ by
\begin{equation}
\bm{L}^*(t)=\bm{L}_0+(\rho^T-\lambda^T)\Pi'\bm{v}^*t
\end{equation}
and by defining $\bm{\ell}^*(t)$ as the solution of $\bm{F}[\bm{\ell}^*(t),\bm{L}^*(t)]=0$. This set manifestly satisfies the initial condition $[\bm{\ell}_0,\bm{L}_0]$ chosen above. Inserting $[\bm{\ell}^*(t),\bm{L}^*(t)]$ into Eqs.~(\ref{eq:diffsystem}), we further verify that these functions satisfies the equations of motion, implying that they describe the dynamics of the bundle. Using Eqs.~(\ref{eq:fv}) and (\ref{eq:v''eq}), we find that $\bm{v}''=0$ and $\rho\bm{v}^l=\lambda\bm{v}^r=\Pi'\bm{v}^*$ for all times, thus confirming that linker units are immobile and that the velocities associated with right- and left-pointing actin are  $v^*$ and $-v^*$, respectively.

In this regime, the maximum relative speed between any two actin filament units is $2|v^*|$ irrespective of bundle geometry. Defining the contraction velocity of the bundle as the difference between the velocity of its leftmost and rightmost filaments, this implies that the contraction velocity cannot exceed the constant $2|v^*|$. It is thus impossible for the contraction velocity to scale linearly with bundle length, and telescopic contractility does not occur. Instead, filaments are segregated according to polarity, as observed experimentally in Ref.~\cite{Tanaka-Takiguchi:2004}. Note that this result does not require the bundle to be disordered.

\subsection{\label{sec:Disordered bundles with linearly elastic filaments}Disordered bundles with linearly elastic filaments}
In a bundle with arbitrary force-velocity relationships, the reasoning of the previous section does not apply, and
some amount of contraction or extension is generally present.
We thus ask whether bundles contract \emph{on average}, and find that they do only if filament polarities are organized across the bundles, or if the filaments display nonlinear elastic behavior.

To prove this statement, we first give a mathematical description of bundles devoid of both polarity organization and filament nonlinear elastic behavior (Sec.~\ref{sec:Mathematical formulation}), and then use Eqs.~(\ref{eq:diffsystem}) and (\ref{eq:v''eq}) to show that such bundles do not contract  (Sec.~\ref{sec:Proof of the property}).

\subsubsection{\label{sec:Mathematical formulation}Mathematical formulation}
Consider an arbitrary bundle, which we denote by $B$. Bundle $B$ is fully characterized by specifying matrices $\lambda$, $\rho$, $\Pi$, $\gamma$, the force-velocity functions $\bm{V}'$ and the initial condition $\bm{L}_0$. Therefore, a population of bundles is fully characterized by specifying the distribution ${\cal P}(B)={\cal P}(\lambda,\rho,\Pi,\gamma,\bm{V}',\bm{L}_0)$ of the frequencies at which any possible bundle $B$ arises in the population. For a population without any polarization organization, ${\cal P}(B)$ must be independent of $\Pi$, implying in particular that it is invariant under polarity reversal:
\begin{equation}\label{eq:nopolorga}
{\cal P}(\lambda,\rho,\Pi,\gamma,\bm{V}',\bm{L}_0)={\cal P}(\lambda,\rho,-\Pi,\gamma,\bm{V}',\bm{L}_0).
\end{equation}
This relationship clearly does not apply to a population of sarcomeres. Indeed, in a sarcomere static cross-linkers are restricted to the barbed ends of F-actin while active, mobile motors are found at the pointed ends. Thus the polarity-reversed image of a sarcomere is not a sarcomere; actually, inverting the polarities of filaments in Fig.~\ref{fig:muscle}(a) results in an extensile, not contractile, structure. Assuming in the following that Eq.~(\ref{eq:nopolorga}) holds thus excludes sarcomeric contractility from our discussion.

We further assume that filament units exhibit linear elastic behavior, which reads
\begin{subequations}\label{eq:linelastic}
\begin{equation}\label{eq:linforceext}
\bm{F}(\bm{\ell},\bm{L})=
\frac{\partial\bm{F}}{\partial\bm{\ell}}(\bm{\ell}-\bm{\ell}_0)
+
\frac{\partial\bm{F}}{\partial\bm{L}}(\bm{L}-\bm{L}_0),
\end{equation}
where ${\partial{F}}/{\partial{\ell}}$ and ${\partial{F}}/{\partial{L}}$ are constants, \emph{i.e.}
\begin{equation}\label{eq:constantspringconst}
\frac{\d}{\d t}\left(\frac{\partial\bm{F}}{\partial\bm{\ell}}\right)
=
\frac{\d}{\d t}\left(\frac{\partial\bm{F}}{\partial\bm{L}}\right)
=0.
\end{equation}
\end{subequations}
This assumption is a good description of very stiff filaments, where $\ell_i=L_i$ for any filament unit $i$. Indeed, such filaments can be described by choosing the force-extension relationship
\begin{equation}\label{eq:Kfext}
F(\ell_i,L_i)=K(\ell_i-L_i),
\end{equation}
which satisfies Eqs.~(\ref{eq:linelastic}) provided that $K$ is a constant, and enforcing the limit $K\rightarrow+\infty$.

\subsubsection{\label{sec:Proof of the property}Proof of the property}
\begin{figure}[t]
\resizebox{8.5cm}{!}{\includegraphics{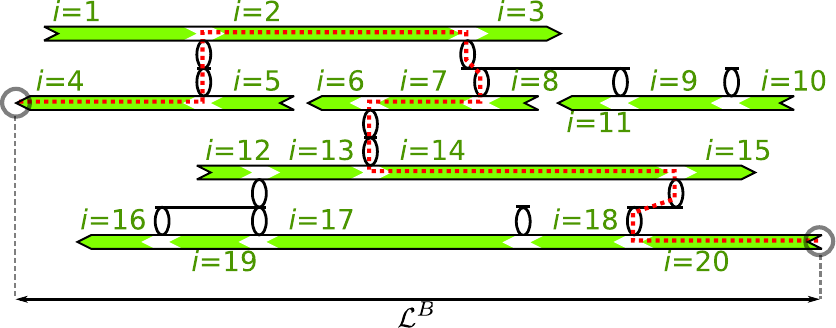}}
\caption{\label{fig:path}
(Color online) Example of a path (\emph{red dotted line}), as discussed in Sec.~\ref{sec:Disordered bundles with linearly elastic filaments}. The leftmost and rightmost points of the bundle are labelled by \emph{gray circles}. In this example $\epsilon_2^B=\epsilon_4^B=\epsilon_{14}^B=\epsilon_{20}^B=-\epsilon_{7}^B=1$ and all other $\epsilon_i^B$s are equal to zero. As a consequence, Eq.~(\ref{eq:totallengthderivative}) reads ${\d {\cal L}^B}/{\d t}={\d}(\ell_4^B+\ell_2^B-\ell_7^B+\ell_{14}^B+\ell_{20}^B)/{\d t}$.
}
\end{figure}

To determine whether a bundle $B=\lbrace\lambda,\rho,\Pi,\gamma,\bm{V}',\bm{L}_0\rbrace$ is contractile or extensile, we imagine labeling its leftmost and rightmost points, and ask whether the distance ${\cal L}^B$ between these two labels tends to increase or decrease with time. To calculate ${\cal L}^B$, we choose a path along the bundle's linker and filament units going from the left label to the right label as pictured in Fig.~\ref{fig:path}. We define $\epsilon_i^B$ as equal to $1$ if the path considered crosses filament unit $i$ from left to right, to $-1$ if it crosses it from right to left, and to $0$ otherwise (see the caption of Fig.~\ref{fig:path}). The contraction velocity of the bundle can then be defined as:
\begin{equation}\label{eq:totallengthderivative}
\frac{\d{\cal L}^B}{\d t}=\sum_{i=1}^{n}\epsilon_i^B\frac{\d \ell_i^B}{\d t}.
\end{equation}
We use the following notation to refer to the solution of the equations of motion for bundle $B$:
\begin{subequations}\label{eq:Delta}
\begin{eqnarray}
{\bm{L}}^B(t)&=&\bm{L}_0+\Delta{\bm{L}}^B(t)\\
{\bm{\ell}}^B(t)&=&\bm{\ell}_0+\Delta{\bm{\ell}}^B(t).
\end{eqnarray}
\end{subequations}

We now introduce bundle $\tilde{B}$ as the polarity-reversed image of $B$, \emph{i.e.}, $\tilde{B}=\lbrace\lambda,\rho,-\Pi,\gamma,\bm{V}',\bm{L}_0\rbrace$. Substituting Eqs.~(\ref{eq:linelastic}) into Eqs.~(\ref{eq:diffsystem}) and (\ref{eq:v''eq}), we find that the dynamics of $\tilde{B}$ satisfies
\begin{subequations}\label{eq:Deltatilde}
\begin{eqnarray}
\Delta{\bm{L}}^{\tilde{B}}(t)=-\Delta\bm{L}^B(t)\\
\Delta{\bm{\ell}}^{\tilde{B}}(t)=-\Delta\bm{\ell}^B(t).
\end{eqnarray}
\end{subequations}
Combining this with Eqs.~(\ref{eq:totallengthderivative}) and (\ref{eq:Delta}) while using the path $\epsilon_i^{\tilde{B}}=\epsilon_i^{{B}}$ to assess the contraction of $\tilde{B}$, we find
\begin{equation}\label{eq:Btildecontract}
\frac{\d{\cal L}^{\tilde{B}}}{\d t}=-\frac{\d{\cal L}^B}{\d t}.
\end{equation}

We finally calculate the average contraction velocity over a population of bundles as
\begin{equation}
\left\langle\frac{\d{\cal L}}{\d t}\right\rangle=\sum_B{\cal P}(B)\frac{\d{\cal L}^B}{\d t},
\end{equation}
where the sum runs over all possible bundles. Reorganizing this sum, we find
\begin{eqnarray}
\left\langle\frac{\d{\cal L}}{\d t}\right\rangle&=&\frac{1}{2}\sum_B\left[{\cal P}(B)\frac{\d{\cal L}^B}{\d t}+{\cal P}(\tilde{B})\frac{\d{\cal L}^{\tilde{B}}}{\d t}\right]\nonumber\\
&=&\frac{1}{2}\sum_B{\cal P}(B)\left(\frac{\d{\cal L}^B}{\d t}+\frac{\d{\cal L}^{\tilde{B}}}{\d t}\right)=0\label{eq:noavgcontract},
\end{eqnarray}
where Eqs.~(\ref{eq:nopolorga}) and (\ref{eq:Btildecontract}) are used to derive the second and third equalities, respectively. Eq.~(\ref{eq:noavgcontract}) demonstrates that bundles without polarity organization or nonlinear elastic behavior do not contract or extend on average. This result does not depend on bundle structure or the form of the motor force-velocity relationships, and can easily be generalized to bundles pinned to a rigid substrate, or to include friction of the linker or filament units with the solvent. Mean-field modeling of dilute actomyosin gels with rigid filaments suggest that this symmetry-based reasoning could have a three-dimensional counterpart \cite{Liverpool:2005}. However, geometrical nonlinearities in two or more dimensions can take on the role played by elastic nonlinearities in one-dimensional bundles, thus enabling contraction in disordered networks of rigid filaments \cite{Dasanayake:2011}.

\section{\label{sec:Discussion}Discussion}
In this paper we consider bundles of sliding motors and filaments with arbitrary geometries and motor force-velocity relationships and show that their contractility requires
\begin{enumerate}
\item non-identical motors
\item and
	\vspace{-2mm}
	\begin{enumerate}
	\item either polarity organization
	\item or nonlinear elastic response of the filaments.
	\end{enumerate}
\end{enumerate}
While our model is framed in term of F-actin and myosin for clarity, our results are much more general and could equally apply to bundles comprised of other types of motors and filaments (\emph{e.g.}, kinesins and microtubules).
Our description includes  as a special case the well-understood contractility of striated muscle sarcomeres [Fig.~\ref{fig:muscle}(a)]. Their architecture includes both identical myosin thick filaments and passive cross-linkers (which are mathematically equivalent to motors with velocity zero), thus satisfying condition (1). They moreover have a distinctive polarity organization, and thereby fulfill condition (2a). Similarly, bundles with motors whose velocities depend on their position relative to the filaments \cite{Kruse:2000,Kruse:2003aa,Liverpool:2003,Ziebert:2005,Liverpool:2005} generically break polarity-reversal symmetry, which results in polarity organization.

Besides establishing the requirements for contractility, the formalism presented here can describe the dynamics of a wide range of contractile bundles. For instance, straightforward numerical simulations of Eqs.~(\ref{eq:diffsystem}) and (\ref{eq:v''eq}) could be used to describe bundle deformation as a function of the initial arrangement of the filament and motors. Comparing these predictions to experimental observations while varying these initial parameters could yield insight into the architecture of the bundles, which is currently lacking. Although such a study is beyond the scope of our work, a simplified version of our formalism still successfully predicts the onset of their contraction \cite{Lenz:2012}.

While the model used here is designed to describe a large class of contractile bundles, some of our assumptions are especially appropriate to describe the reconstituted bundles of Ref.~\cite{Thoresen:2011}. F-actin is phalloidin-stabilized in this system, implying that no actin polymerization-depolymerization takes place; myosin thick filaments are much shorter ($\simeq 300\,$nm) and thicker ($\simeq 50\,$nm) than F-actin ($\simeq 5\,\mu$m and $\simeq 5\,$nm, respectively), justifying our assumption that they behave as rigid objects; and myosin thick filaments do not detach from the bundle on the time scales relevant for contraction, suggesting that the filament-motor attachment-detachment dynamics are inessential to contractility. Indeed, we show elsewhere that attachment-detachment can limit contractility under low myosin conditions \cite{Lenz:2012}.

Going beyond the specifics of the system studied in Ref.~\cite{Thoresen:2011}, it is interesting to discuss bundles where this attachment-detachment dynamics is not negligible. In the simplest such situation, motors undergo attachment and detachment at a constant rate, which tends to randomize their distribution in a filament polarity-independent manner. This fails to break the polarity-reversal symmetry discussed in this paper, and the resulting requirements for contractility are unchanged.

\begin{figure}[t]
\resizebox{8.5cm}{!}{\includegraphics{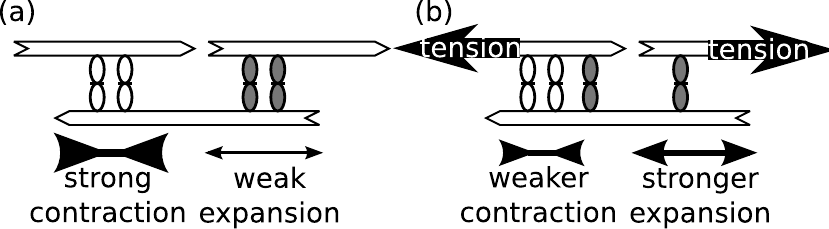}}
\caption{\label{fig:fenn}
The Fenn effect impedes bundle contractility.
(a)~Simple contractile bundle comprised of fast (\emph{light}) and slow (\emph{dark}) motors. \emph{Double-headed arrows} indicate local contraction or expansion due to the movement of the motors relative to the filaments [Fig.~\ref{fig:muscle}(b)].
(b)~As tension builds following contraction, the motors detach and reorganize and contractility diminishes.
}
\end{figure}

A more interesting question is to ask whether contractility could arise from the load-dependent detachment of myosin motors. Specifically, the detachment rate of myosin motors decreases under increasing load, a tendency known as the ``Fenn effect''  \cite{Veigel:2003}. To assess its influence on contractility, we consider the simple bundle of rigid filaments illustrated in Fig.~\ref{fig:fenn}(a). This bundle comprises a contracting and an expanding region similar to those of Fig.~\ref{fig:muscle}(b). When faster motors happen to be concentrated in the contracting region, this bundle is contractile. Assuming that its ends are fixed, the bundle comes under tensile force during contraction. As a consequence, its expanding region comes under negative load and the motors there tend to detach. Conversely, the motors in the contracting region experience a positive load and tend to hold on to the filaments. On average, detached motors thus tend to diffuse from the expanding to the contracting region [Fig.~\ref{fig:fenn}(b)]. The reduced number of motors in the expanding region implies that its resistance to the applied tension decreases, thus increasing its expansion rate. As the contracting region tends to recruit the slower motors from the expanding region, its contraction rate decreases. Overall, the Fenn effect thus tends to suppress contractile configurations rather than amplify them. This suggests that the Fenn effect cannot generate contractility in the absence of polarity organization or filament nonlinear elastic response.

\begin{figure}[t]
\resizebox{8.5cm}{!}{\includegraphics{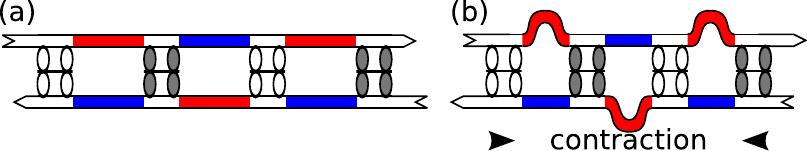}}
\caption{\label{fig:buckle}
Filament buckling as a mechanism for bundle contractility.
(a)~The presence of fast (\emph{white}) and slow (\emph{gray}) motors generically induce compressive (\emph{red}) and extensile (\emph{blue}) stresses in filaments.
(b)~Buckling of the compressed filaments leads to an overall shortening of the bundle.
Buckling of filament units of length $L_i=500\,$nm occurs for compressive forces of order $k_BT\ell_p/L_i^2\simeq 0.16\,$pN, where $\ell_p\simeq 10\,\mu$m is the F-actin persistence length. Myosin thick filaments typically exert forces of several piconewtons \cite{Molloy:1995}, which is sufficient to induce buckling, but remains smaller than the tension required to break F-actin ($\simeq 100\,$pN \cite{Yanagida:1988}).
}
\end{figure}

The requirements derived in this paper offer insight into the contractility of actomyosin bundles devoid of sarcomere-like organization, for which no established contraction mechanism exists. As they do not satisfy condition (2a), we propose that they contract by fulfilling conditions (1) and (2b) \cite{Lenz:2012}. Consider two antiparallel filaments interacting through several different motors with distinct speeds [Fig.~\ref{fig:buckle}(a)]. As motors start to move relative to the filaments, stresses build in sections of the filament flanked by motors with different speeds. When the flanking motor proximal to the barbed end is faster than that proximal to the pointed end, compression arises. When it is slower, tension arises. F-actin responds nonlinearly to these stresses by buckling under compression while resisting extension,
which has previously been proposed to play a role in actomyosin contraction \cite{Silva:2011}.
Following buckling of the compressed filament sections, fast motors are free to move quickly while the others move slowly. This results in the growth of the compressed sections and shrinkage of the extended ones, and thus in overall bundle contraction [Fig.~\ref{fig:buckle}(b)]. Experimental observations suggest that this mechanism could be at the origin of contraction in reconstituted actomyosin bundles \cite{Lenz:2012},
and that F-actin buckling can occur in cells \cite{Costa:2002}. These results offer
an interesting new perspective on mechanisms underlying actomyosin bundle contraction \emph{in vivo}.

\begin{acknowledgments}
We thank Yitzhak Rabin, Todd Thoresen and Tom Witten for inspiring discussions. This work was supported by NSF DMR-MRSEC 0820054
\end{acknowledgments}


\begin{thebibliography}{10}

\bibitem{Alberts:1998aa}
Bruce Alberts, Dennis Bray, Alexander Johnson, Julian Lewis, Martin Raff, Keith
  Roberts, and Peter Walter.
\newblock {\em Essential Cell Biology}.
\newblock Garland, New-York, 1998.

\bibitem{Peterson:2004}
Lynda~J. Peterson, Zenon Rajfur, Amy~S. Maddox, Christopher~D. Freel, Yun Chen,
  Magnus Edlund, Carol Otey, and Keith Burridge.
\newblock Simultaneous stretching and contraction of stress fibers in vivo.
\newblock {\em Mol. Biol. Cell}, 15(7):3497--3508, July 2004.

\bibitem{Fay:1983}
F.~S. Fay, K.~Fujiwara, D.~D. Rees, and K.~E. Fogarty.
\newblock Distribution of alpha-actinin in single isolated smooth muscle cells.
\newblock {\em J. Cell Biol.}, 96(3):783--795, March 1983.

\bibitem{Heath:1983}
J.~P. Heath.
\newblock Behaviour and structure of the leading lamella in moving fibroblasts.
  {I}. occurrence and centripetal movement of arc-shaped microfilament bundles
  beneath the dorsal cell surface.
\newblock {\em J. Cell Sci.}, 60:331--354, March 1983.

\bibitem{Cramer:1997}
L.~P. Cramer, M.~Siebert, and T.~J. Mitchison.
\newblock Identification of novel graded polarity actin filament bundles in
  locomoting heart fibroblasts: implications for the generation of motile
  force.
\newblock {\em J. Cell Biol.}, 136(6):1287--1305, March 1997.

\bibitem{Medalia:2002}
Ohad Medalia, Igor Weber, Achilleas~S Frangakis, Daniela Nicastro, Gunther
  Gerisch, and Wolfgang Baumeister.
\newblock Macromolecular architecture in eukaryotic cells visualized by
  cryoelectron tomography.
\newblock {\em Science}, 298(5596):1209--1213, November 2002.

\bibitem{Verkhovsky:1995}
A.~B. Verkhovsky, T.~M. Svitkina, and G.~G. Borisy.
\newblock Myosin {II} filament assemblies in the active lamella of fibroblasts:
  their morphogenesis and role in the formation of actin filament bundles.
\newblock {\em J. Cell Biol.}, 131(4):989--1002, November 1995.

\bibitem{Szent-Gyorgyi:1947}
A.~Szent-Gy\"orgyi.
\newblock {\em Chemistry of muscular contraction}.
\newblock Academic Press, New York, 1947.

\bibitem{Janson:1991}
L.~W. Janson, J.~Kolega, and D.~L. Taylor.
\newblock Modulation of contraction by gelation/solation in a reconstituted
  motile model.
\newblock {\em J. Cell Biol.}, 114(5):1005--1015, September 1991.

\bibitem{Mizuno:2007aa}
Daisuke Mizuno, Catherine Tardin, C.~F. Schmidt, and F.~C. Mackintosh.
\newblock Nonequilibrium mechanics of active cytoskeletal networks.
\newblock {\em Science}, 315(5810):370--373, January 2007.

\bibitem{Bendix:2008}
Poul~M. Bendix, Gijsje~H. Koenderink, Damien Cuvelier, Zvonimir Dogic, Bernard~N.
  Koeleman, William~M. Brieher, Christine~M. Field, L.~Mahadevan, and David~A.
  Weitz.
\newblock A quantitative analysis of contractility in active cytoskeletal
  protein networks.
\newblock {\em Biophys. J.}, 94(8):3126--36, April 2008.

\bibitem{Koenderink:2009}
Gijsje~H. Koenderink, Zvonimir Dogic, Fumihiko Nakamura, Poul~M. Bendix,
  Frederick~C. MacKintosh, John~H. Hartwig, Thomas~P. Stossel, and David~A. Weitz.
\newblock An active biopolymer network controlled by molecular motors.
\newblock {\em Proc. Natl. Acad. Sci. U.S.A.}, 106(36):15192--7, September
  2009.

\bibitem{Thoresen:2011}
Todd Thoresen, Martin Lenz, and Margaret~L. Gardel.
\newblock Reconstitution of contractile actomyosin bundles.
\newblock {\em Biophys. J.}, 100(11):2698--2705, June 2011.

\bibitem{Kruse:2003ab}
K.~Kruse, A.~Zumdieck, and F.~J\"ulicher.
\newblock Continuum theory of contractile fibres.
\newblock {\em Europhys. Lett.}, 64(5):716--722, December 2003.

\bibitem{Kruse:2004aa}
K.~Kruse, J.-F. Joanny, F.~J{\"u}licher, J.~Prost, and K.~Sekimoto.
\newblock Asters, vortices, and rotating spirals in active gels of polar
  filaments.
\newblock {\em Phys. Rev. Lett.}, 92:078101, January 2004.

\bibitem{Kruse:2005aa}
K.~Kruse, J.~F. Joanny, F.~J{\"u}licher, J.~Prost, and K.~Sekimoto.
\newblock Generic theory of active polar gels: a paradigm for cytoskeletal
  dynamics.
\newblock {\em Eur. Phys. J. E}, 16(1):5--16, January 2005.

\bibitem{MacKintosh:2008aa}
F.~C. MacKintosh and A.~J. Levine.
\newblock Nonequilibrium mechanics and dynamics of motor-activated gels.
\newblock {\em Phys. Rev. Lett.}, 100(1):018104, January 2008.

\bibitem{Zemel:2009}
Assaf Zemel and Alex Mogilner.
\newblock Motor-induced sliding of microtubule and actin bundles.
\newblock {\em Phys. Chem. Chem. Phys.}, 11(24):4821--4833, June 2009.

\bibitem{Kruse:2000}
K.~Kruse and F.~J{\"u}licher.
\newblock Actively contracting bundles of polar filaments.
\newblock {\em Phys. Rev. Lett.}, 85(8):1778--1781, August 2000.

\bibitem{Kruse:2003aa}
Karsten Kruse and Franck J\"ulicher.
\newblock Self-organization and mechanical properties of active filament
  bundles.
\newblock {\em Phys. Rev. E}, 67(5):051913, May 2003.

\bibitem{Liverpool:2003}
Tanniemola~B. Liverpool and M.~Cristina Marchetti.
\newblock Instabilities of isotropic solutions of active polar filaments.
\newblock {\em Phys. Rev. Lett.}, 90(13):138102, April 2003.

\bibitem{Ziebert:2005}
F.~Ziebert and W.~Zimmermann.
\newblock Nonlinear competition between asters and stripes in filament-motor
  systems.
\newblock {\em Eur. Phys. J. E}, 18(1):41--54, September 2005.

\bibitem{Liverpool:2005}
T.~B. Liverpool and M.~C. Marchetti.
\newblock Bridging the microscopic and the hydrodynamic in active filament
  solutions.
\newblock {\em Europhys. Lett.}, 69(5):846--852, March 2005.

\bibitem{Note1}
Mechanical stability imposes that ${\partial {F}}/{\partial {\ell }}$ is always
  strictly negative, making the matrix ${\partial \protect \bm {F}}/{\partial
  \protect \bm {\ell }}$ invertible.

\bibitem{Bement:1991}
W.~M. Bement and D.~G. Capco.
\newblock Analysis of inducible contractile rings suggests a role for protein
  kinase {C} in embryonic cytokinesis and wound healing.
\newblock {\em Cell Motil. Cytoskeleton}, 20(2):145--157, 1991.

\bibitem{Herrera:2005}
Ana~M. Herrera, Brent~E. McParland, Agnes Bienkowska, Ross Tait, Peter~D.
  Par{\'e}, and Chun~Y. Seow.
\newblock ``sarcomeres'' of smooth muscle: functional characteristics and
  ultrastructural evidence.
\newblock {\em J. Cell Sci.}, 118(11):2381--2392, June 2005.

\bibitem{Carvalho:2009}
Ana Carvalho, Arshad Desai, and Karen Oegema.
\newblock Structural memory in the contractile ring makes the duration of
  cytokinesis independent of cell size.
\newblock {\em Cell}, 137(5):926--937, May 2009.

\bibitem{Tanaka-Takiguchi:2004}
Yohko Tanaka-Takiguchi, Toshihito Kakei, Akinori Tanimura, Aya Takagi, Makoto
  Honda, Hirokazu Hotani, and Kingo Takiguchi.
\newblock The elongation and contraction of actin bundles are induced by
  double-headed myosins in a motor concentration-dependent manner.
\newblock {\em J. Mol. Biol.}, 341(2):467--76, August 2004.

\bibitem{Dasanayake:2011}
Nilushi~L Dasanayake, Paul~J Michalski, and Anders~E. Carlsson.
\newblock General mechanism of actomyosin contractility.
\newblock {\em Phys. Rev. Lett.}, 107(11):118101, September 2011.

\bibitem{Lenz:2012}
Martin Lenz, Todd Thoresen, Margaret~L. Gardel, and Aaron~R. Dinner.
\newblock Contractile units in disordered actomyosin bundles arise from
  {F}-actin buckling.
\newblock {ar{X}iv}:1201.4110, January 2012.

\bibitem{Veigel:2003}
Claudia Veigel, Justin~E. Molloy, Stephan Schmitz, and John Kendrick-Jones.
\newblock Load-dependent kinetics of force production by smooth muscle myosin
  measured with optical tweezers.
\newblock {\em Nat. Cell Biol.}, 5(11):980--986, November 2003.

\bibitem{Silva:2011}
Marina Soares~{e} Silva, Martin Depken, Bj{\"o}rn Stuhrmann, Marijn Korsten,
  Fred~C. Mackintosh, and Gijsje~H. Koenderink.
\newblock Active multistage coarsening of actin networks driven by myosin
  motors.
\newblock {\em Proc. Natl. Acad. Sci. U.S.A.}, 108(23):9408--9413, June 2011.

\bibitem{Costa:2002}
Kevin~D. Costa, William~J. Hucker, and Frank C.-P. Yin.
\newblock Buckling of actin stress fibers: a new wrinkle in the cytoskeletal
  tapestry.
\newblock {\em Cell Motil. Cytoskeleton}, 52(4):266--274, August 2002.

\bibitem{Molloy:1995}
J.~E. Molloy, J.~E. Burns, J.~Kendrick-Jones, R.~T. Tregear, and D.~C.~S.
  White.
\newblock Movement and force produced by a single myosin head.
\newblock {\em Nature}, 378(6553):209--212, November 1995.

\bibitem{Yanagida:1988}
Akiyoshi Kishino and Toshio Yanagida.
\newblock Force measurements by micromanipulation of a single actin filament by
  glass needles.
\newblock {\em Nature}, 334(6177):74--76, July 1988.

\end{thebibliography}
\end{document}